\documentclass[12pt,preprint]{aastex}
\shorttitle{The Radial Velocity Variability of $\gamma$~Dra}
\shortauthors{Hatzes et al.}
\begin{document}

   \title{The Radial Velocity Variability of the K-giant $\gamma$~Draconis: \\
Stellar Variability Masquerading as a Planet}

\author{A. P. Hatzes}
\affil{Th\"uringen  Landessternwarte Tautenburg, Sternwarte 5, D-07778 Tautenburg, Germany}
\email{artie@tls-tautenburg.de}
\and
\author{M. Endl}
\affil{McDonald Observatory, The University of Texas at Austin,
    Austin, TX 78712, USA}
\and
\author{W. D. Cochran}
\affil{McDonald Observatory, The University of Texas at Austin,
    Austin, TX 78712, USA}
\and
 \author{P. J. MacQueen}
\affil{McDonald Observatory, The University of Texas at Austin,
    Austin, TX 78712, USA}
\and
 \author{I. Han}
 \affil{Korea Astronomy and Space Science Institute, 776 daedukdae-ro, Yuseong-gu,
305-348, Taejeon, South Korea}
 \and
\author{B.-C. Lee}
 \affil{Korea Astronomy and Space Science Institute, 776 daedukdae-ro, Yuseong-gu,
305-348, Taejeon, South Korea}
\and
\author{K.-M. Kim}
 \affil{Korea Astronomy and Space Science Institute, 776 daedukdae-ro, Yuseong-gu,
305-348, Taejeon, South Korea}
\and 
\author{D. Mkrtichian}
\affil{National Astronomical Research Institute of Thailand, 191
Siriphanich Bldg., Huay Kaew Rd., Suthap, Muang, 50200
Chiang Mai, Thailand}
\affil{Astronomical Observatory,  Odessa National University, Schechenko Park, Odessa,
65014, Ukraine}

\and
\author{M. D\"ollinger}
\affil{Th\"uringen  Landessternwarte Tautenburg, Sternwarte 5, D-07778 Tautenburg, Germany}
\and
\author{M. Hartmann}
\affil{Th\"uringen  Landessternwarte Tautenburg, Sternwarte 5, D-07778 Tautenburg, Germany}
\and
\author{M. Karjalainen}
\affil{Isaac Newton Group of Telescopes, Apartado de Correos 321, Santa Cruz de la Palma, E-38700, Spain}
\and
\author{S. Dreizler}
\affil{Institut f\"ur Astrophysik, Georg-August-Universit\"at, Friedrich-Hund-Platz, 1, 37077, G\"ottingen,
Germany}

\begin{abstract}
We present precise stellar radial velocity measurements of 
$\gamma$~Dra taken from 2003 to 2017.
The data from 2003 to  2011 show coherent, long-lived variations with a period
of 702 d.
These variations are consistent  with the presence of a planetary companion
having $m$ sin$i$ = 10.7 M$_{\rm{Jup}}$  whose orbital properties are typical for
giant planets found around evolved stars.
An analysis of the {\rm Hipparcos} photometry, Ca II S-index
measurements, and measurements of the spectral line shapes during this time show no variations with the
radial velocity of the planet which seems to ``confirm'' the presence of the planet.
However, radial velocity measurements
taken 2011 -- 2017 seem to refute this.
From 2011 to 2013 the radial velocity variations virtually 
disappear only to return in 2014, but with a noticeable phase shift. The total radial velocity
variations are consistent either with amplitude variations on timescales of  $\approx$ 10.6 yr, or the
beating effect between two periods of 666 d and  801 d.  It seems unlikely that both these
signals stem from a two-planet system. A simple dynamical analysis indicates that there is only a 1--2\%
chance that the  two-planet
is stable.  Rather, we suggest that this multi-periodic behavior may represent a new form
of stellar variability, possibly related to oscillatory convective modes. 
If such intrinsic stellar variability is common
around K giant stars and is
attributed to planetary companions, then the planet occurrence rate 
among these stars may be significantly lower than thought. 
\end{abstract}


\section{Introduction}
Currently, G-K giants offer us one of the few means of studying the frequency 
of planets around intermediate mass (IM) stars  (1.3 -- 3 $M_\odot$) using the Doppler
method.  While on the main sequence an IM star is ill-suited
for radial velocity (RV) measurements because it has  few stellar lines that are often broadened by high rates of rotation. On the other hand,
IM stars that have evolved onto the giant branch are cool, have lots of stellar
lines and low rotation rates. A number of surveys have searched for exoplanets around giant stars with RV measurements
(Setiawan et al. 2003; D\"ollinger et al.
2007; Johnson et al. 2007; Sato et al. 2008; Niedzielski et al. 2009; Wittenmyer et al.
2011; Jones et al. 2011; Lee et al. 2012). 

Recently, Hrudkov\'a et al. (2017) started a program to search for exoplanets around K giant
stars observed by the {\it Kepler} mission. This program exploits the fact
that the stellar oscillations seen in the {\it Kepler} light curves can
be used to derive accurate stellar masses. Normally, one has to rely on theoretical evolutionary
tracks and their inherent uncertainties. This program discovered the planet candidate around HD~175370.
To date about 100 giant planets have been claimed around evolved G-K stars and most of these have orbital periods of several hundreds of days. Many of these are exoplanets whose nature has been confirmed. These include $\iota$ Dra b in an eccentric orbit (Frink et al. 2002)
and more recently the transiting planet Kepler-91b (Barclay et al. 2015) 

Because of the evolved status of K giant stars (i.e. extended atmospheres, deep convection zones, etc.) it should be  a concern that these RV variations may actually arise from intrinsic
stellar variability. Lately, there have been several planet candidates around K giants 
that can best be termed ``problematical''.  Ramm et al. (2009) discovered
periodic variations in the K giant primary component of the binary system $\nu$ Oct. These had a period of 418\,d and an amplitude of 50 m\,s$^{-1}$ which were consistent with a planetary companion having
a mass of a 2.4 
$M_{Jup}$. However, the orbital period of the binary system was only 1050 days which made the
planet-binary system dynamically unstable. The stability of the
system was only possible if the planet had
a retrograde orbit with respect to the binary motion (Eberle \& Cuntz 2010). More
evidence supporting
this hypothesis came
from a lack of changes in the spectral line shapes as  well as no observed line-depth variations, which were a measure
of temperature variations, with the orbital period (Ramm 2015). It seems that the RV variations were not due
to intrinsic stellar variability.

A recent study by Ramm et al. (2016) provided mixed results as to the reality of $\nu$ Oct Ab.
Additional RV measurements for the star placed precise orbital and mass constraints on the retrograde
planet, but these models were not stable for 10$^6$ years. However, the study did reveal a narrow
range of parameter space where a few orbital models were stable for more than 10$^8$ years.
Although the planet hypothesis seems less likely from a dynamical point of view, 
there is a small chance that the planet
lies on a stable orbit.

There have been other ``dynamically challenged" planets that have been found around other K giant stars. BD+20\,2457 shows RV variations consistent with two planetary candidates with masses  of  21.4 $M_{Jup}$ and 12.5 $M_{Jup}$ and periods of 380 d and 622 d, respectively
(Niedzielski et al. 2009).
No dynamically stable configuration was found for this system. Niedzielski et al. (2009)
found RV variations in HD 102272 that were possibly due to
 two companions with masses of 5.9 $M_{Jup}$ and 2.6 $M_{Jup}$ orbiting
in nearly a 4:1 resonance ($P$ = 127 d and 520 d). However, a dynamical study showed that the planets would quickly collide with one another.  Trifonov et al. (2014) found two giant planets in a 2:1 resonance around $\eta$ Cet whose orbits was only stable for some configurations.
The dynamical problems of all of these systems could be resolved if the RV variations of one of the purported planets were in fact due to stellar variability.

One star in the McDonald sample of K giant stars that was surveyed with
RV measurements by Hatzes
\& Cochran (1993) from over two decades ago was  $\gamma$~Dra
(= HR 6705 = HD 164058 = HIP 87833). RV measurements for this
star showed long-term variability, but there was insufficient data to
determine an accurate  period. The RV monitoring of this star stopped in 1993, but
re-started  in 2003 in order to
investigate further these long-term  variations.
These new measurements showed  RV variations seem to be 
consistent with a planetary companion; however, additional measurements
refute this hypothesis. 
$\gamma$~Dra may be a case of stellar variability in a K giant which can mimic, for a time,
the signal of a planet. 

\section{Stellar Parameters}

The star $\gamma$~Dra has a spectral type of K5 III and is located at a
distance of 45 pc (van Leeuwen, 2007).
Mozurkewich et al. (2003) measured
an angular diameter of 9.86 $\pm$ 0.128 mas. For the distance of $\gamma$~Dra this
implies a stellar radius of 50.03 $\pm$ 0.69 $R_\odot$. 

Several investigations have determined
the  stellar parameters. McWilliam (1990)
measured an effective temperature of 3990 K, a surface gravity of
log $g$ $=$ 1.55, and an abundance that is slightly metal poor,
$[Fe/H]$ $=$ $-$0.14 $\pm$ 0.16. More recent measurements point to a higher
metalicity for the star. Prugniel et al. (2011)
 found $[Fe/H]$ $=$ $+0.11$ $\pm$ 0.05, $T_{\rm eff}$ = 3990 $\pm$ 60 K, and
log $g$ $=$ 1.64 $\pm$ 0.1. These values were consistent
with those derived by Koleva \& Vazdekis (2012): $[Fe/H]$ $=$ $+0.11$ $\pm$ 0.09, 
$T_{\rm eff}$ = 3990 $\pm$ 42 K, and
log $g$ $=$ 1.669 $\pm$ 0.1. 

The basic stellar parameters of mass, radius, and age were determined
using the online tool from Girardi (http://stev.oapd.inaf.it/cgi-bin/param; version PARAM 1.3).
This uses a library of theoretical isochrones (Girardi et al. 2000) and a 
 modified version of the 
Jo$\!\!\!/$rgensen \& Lindegren's (2005) method. A detailed description of this
method is given by da Silva et al. (2006). 

The values of the calculated stellar parameters depend on the measured
input parameters. Using the effective temperature and abundance of
Koleva \& Vazdekis (2012) results in $M$ = 2.14 $\pm$ 0.16 M$_\odot$,
$R$ = 49.07 $\pm$ 3.75 R$_\odot$, and an age of 1.28 $\pm$ 0.29 Gyrs.
The radius determined by this method, $R$ $=$ 49.03 $\pm$ 2.5 R$_\odot$,
 is in excellent agreement with the interferometric value.

The luminosity of the star can be estimated  using two approaches. With the first method 
we combine  the stellar distance, apparent magnitude and bolometric correction to  obtain  the stellar luminosity. The absolute magnitude of $\gamma$~Dra is $-1.144$ $\pm$ 0.015. Buzzoni et al. (2010) 
give a bolometric correction of $-$0.99 $\pm$ 0.10  for $T_{\rm eff}$ = 3990 K. 
This yields a total luminosity of 510  $\pm$ 51 $L_\odot$.
Alternatively,  one can  use the measured stellar radius and effective temperature to calculate the total
luminosity. For $T_{\rm eff}$ = 3990 K this yields $L$ = 515 $\pm$ 37 $L_\odot$.
We simply adopt 510 $L_\odot$ as the luminosity of $\gamma$~Dra.
Table 1 summarizes the stellar parameters.

\section{The Data Sets }

A total of four RV data sets  were used in the analysis. These include
observations made at the 2.1m and 2.7m telescopes at McDonald Observatory,
the 1.8m telescope at the Bohunyanson Optical Astronomy Observatory (BOAO), and the 
2m telescope at the Th\"uringer Landessternwarte Tautenburg (TLS).
Table~2  lists the journal of observations including the time span of the 
measurements and the number of observations. Table~3 lists the RV measurements listed according to
data sets.

\subsection{The McD-2.1 Data Set}

The earliest observations were made with the coude spectrograph of the
2.1m Otto Struve Telescope at McDonald Observatory. A 1200 grooves  mm$^{-1}$ grating 
was used in second   order  in combination with a 
Tektronix 512$\times$ 512 CCD.  Blocking filters were used to isolate the
desired order. This instrumental setup resulted in a  spectral 
dispersion of
0.046\thinspace{\AA} pixel$^{-1}$ at the central wavelength of 
5520\thinspace{\AA}.
An 85 $\mu$m slit provided a spectral resolution
of 0.11\thinspace{\AA} (resolving power $R$ =$\lambda$/$\Delta \lambda$ = 50\,000). 
An iodine absorption cell placed in the light path of the spectrograph
during the stellar observations  provided
the simultaneous wavelength calibration. See Hatzes \& Cochran (1993) for more details of the
instrumental setup and data reductions. Typically 2--10 observations
were made of this star each night.  These data are listed as ``McD-2.1'' in Table 3.

\subsection{The McD-2.7 Data Set}

The ``McD-2.7'' data set used the Tull
 Spectrograph
(Tull et al. 1994) at the Harlan J. Smith 2.7m telescope.
This instrument 
provided a nominal wavelength coverage of
3600\,{\AA} -- 10\,000\,{\AA} at a resolving power of $R$ =  60\,000.  The RV measurement was extracted using self-calibrated
$I_2$ spectra and the 
$Austral$ RV-code (Endl, K\"urster, \& Els 2000).

\subsection{The BOAO Data Set}

RV measurements were also made  with
Bohyunsan Observatory Echelle Spectrograph or BOES (Kim et al. 2006) of
the 1.8m telescope at the BOAO in South Korea. 
A 80  $\mu$m fiber resulted in a resolving power of
$R$ =  90\,000 with a wavelength coverage of 3600 -- 10500\,{\AA}.
Relative RV measurements were  made using an iodine absorption cell as
a reference. These data are referred to 
as the ``BOAO'' set in Table~3.

\subsection{The TLS Data Set}

The TLS observations of $\gamma$~Dra  were made as part of the Tautenburg
Observatory Planet Search (TOPS) program using the 
high-resolution coude echelle spectrometer  of the Alfred Jensch 2m
telescope and an iodine absorption cell placed in the optical
path. The spectrometer is grism cross-dispersed  and it has 
a resolving power $R$ = 67\,000 and 
wavelength coverage 4630--7370\,{\AA}.
A more detailed description of radial velocity measurements
from the TOPS program can be found in Hatzes et al. (2005). 
RV values are listed as ``TLS'' in Table~3.

\begin{figure}[t]
\resizebox{\hsize}{!}{\includegraphics{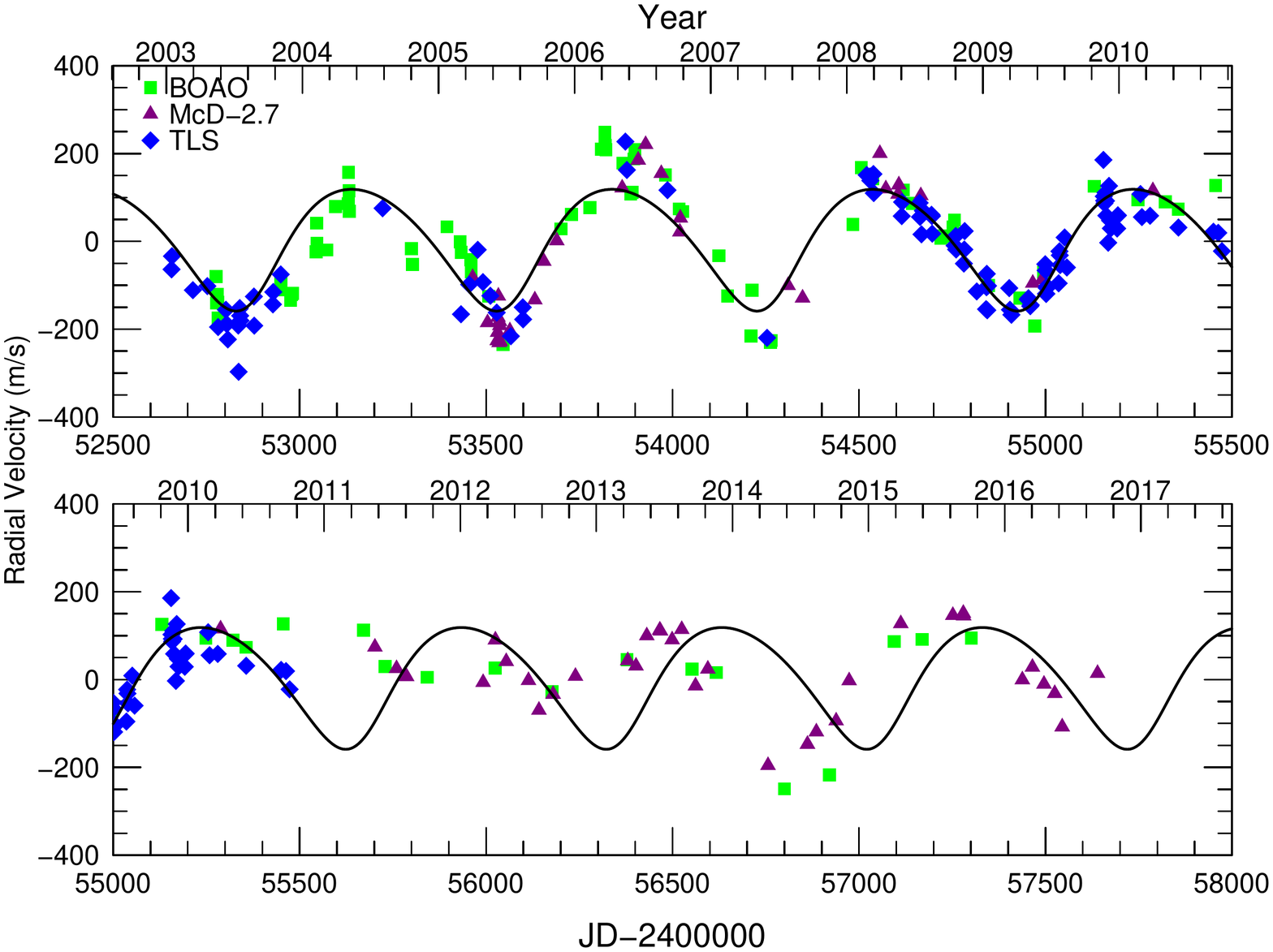}}
\caption{Radial velocity measurements for $\gamma$ Dra from 2003 to 2017
taken with 3 different telescopes and instruments.
Symbols: Triangles :  McD-2.7, diamonds : TLS, squares : BOAO. The solid
line represents the orbital solution (Table 4).
}
\label{orbit}
\end{figure}

\begin{figure}[h]
\resizebox{\hsize}{!}{\includegraphics{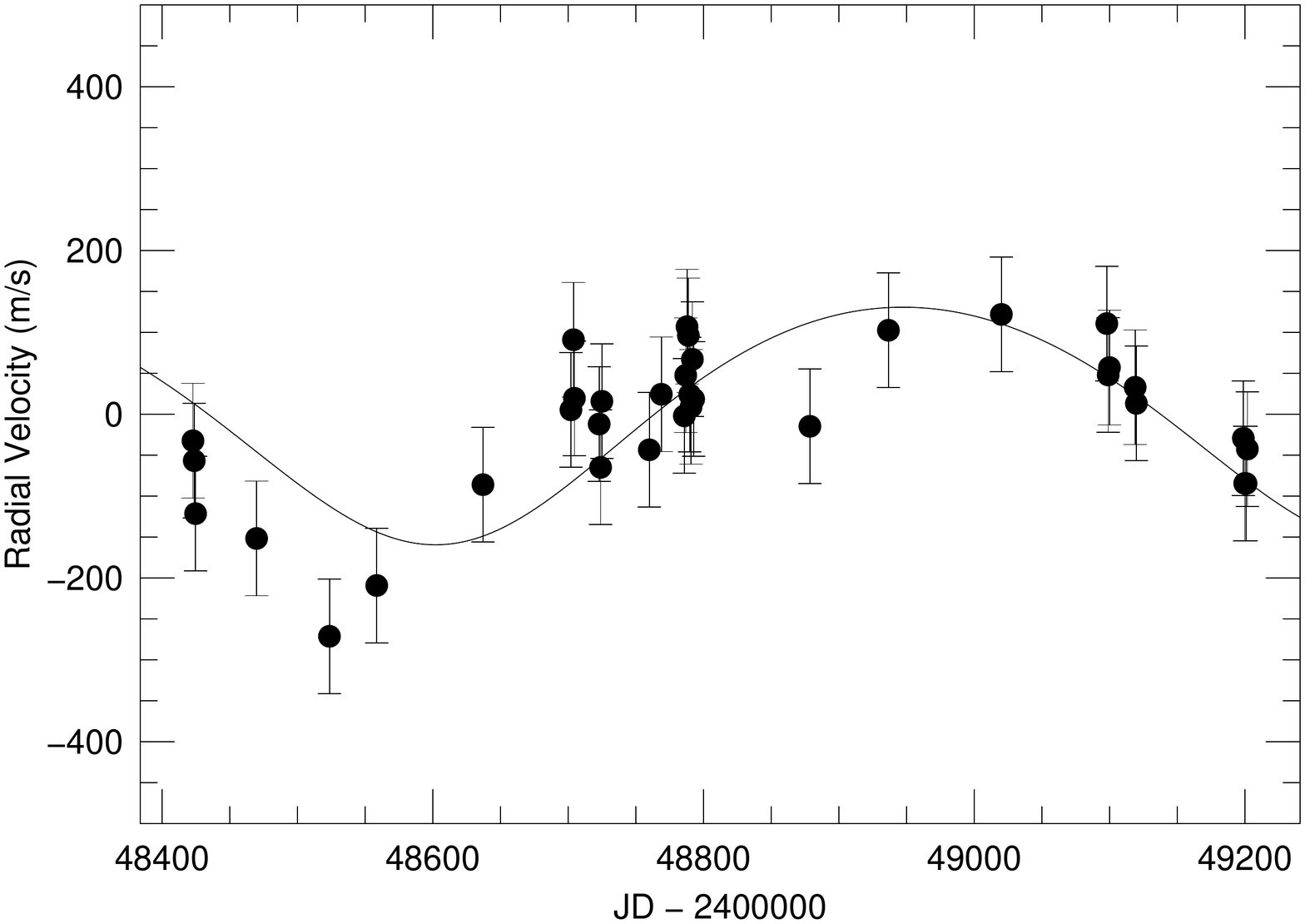}}
\caption{The McD-2.1 RV data (dots) and the orbital solution (line)
using all the RV measurements taken up to 2011. The curve represents
the orbital solution.
}
\label{mcdorbit}
\end{figure}

\begin{figure}[h]
\resizebox{\hsize}{!}{\includegraphics{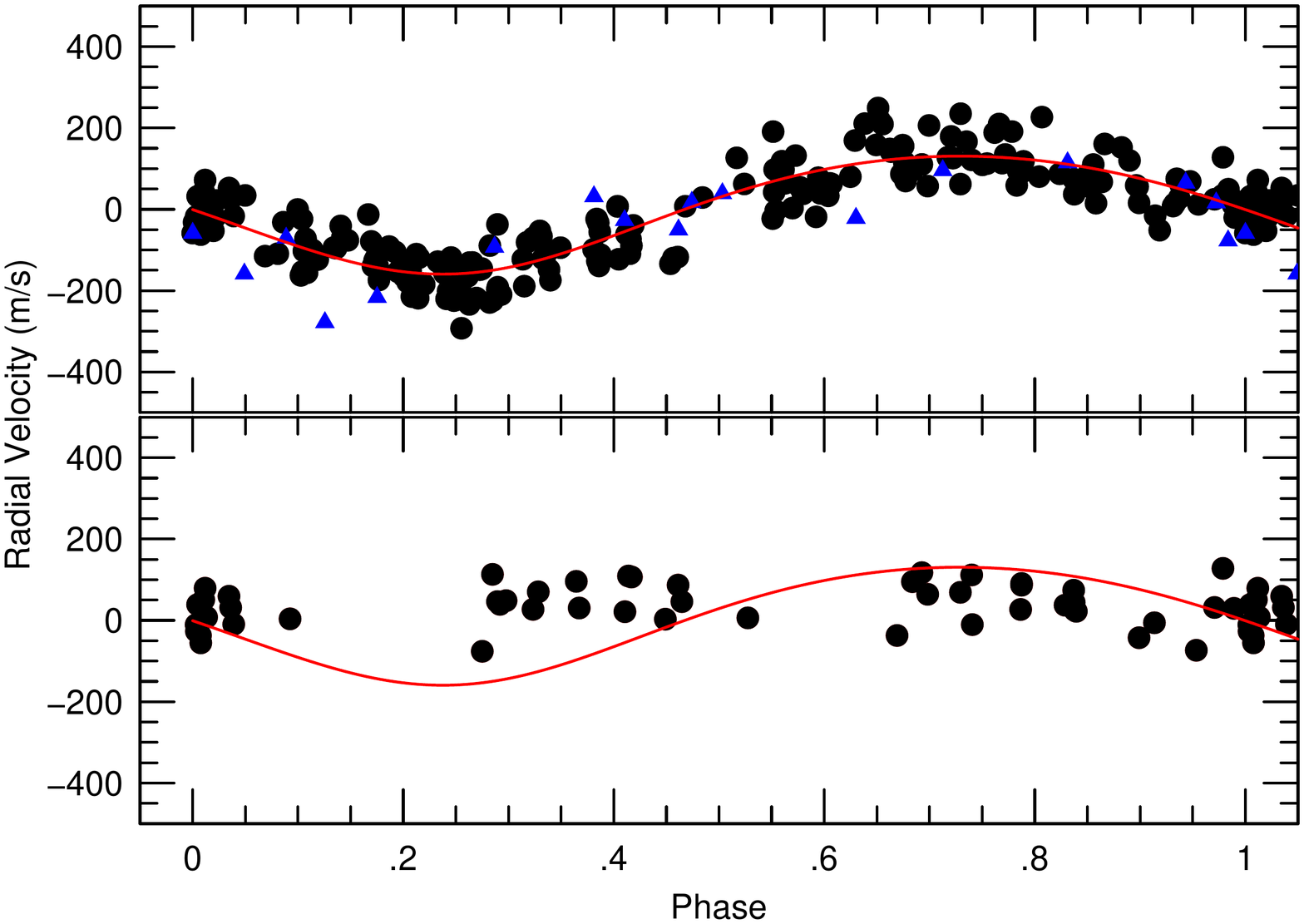}}
\caption{(Top) Radial velocity measurements for $\gamma$~Dra taken
2003-2011 (dots) and phased to the
orbital period of 702 d. The triangles represent the McD-2.1 data taken
20 years earlier.  (Bottom) RV measurements taken 2011-2014 phased to the orbital period.
The curve in both panels represent the orbital solution.
}
\label{phaseorbit}
\end{figure}

\begin{figure}[h]
\resizebox{\hsize}{!}{\includegraphics{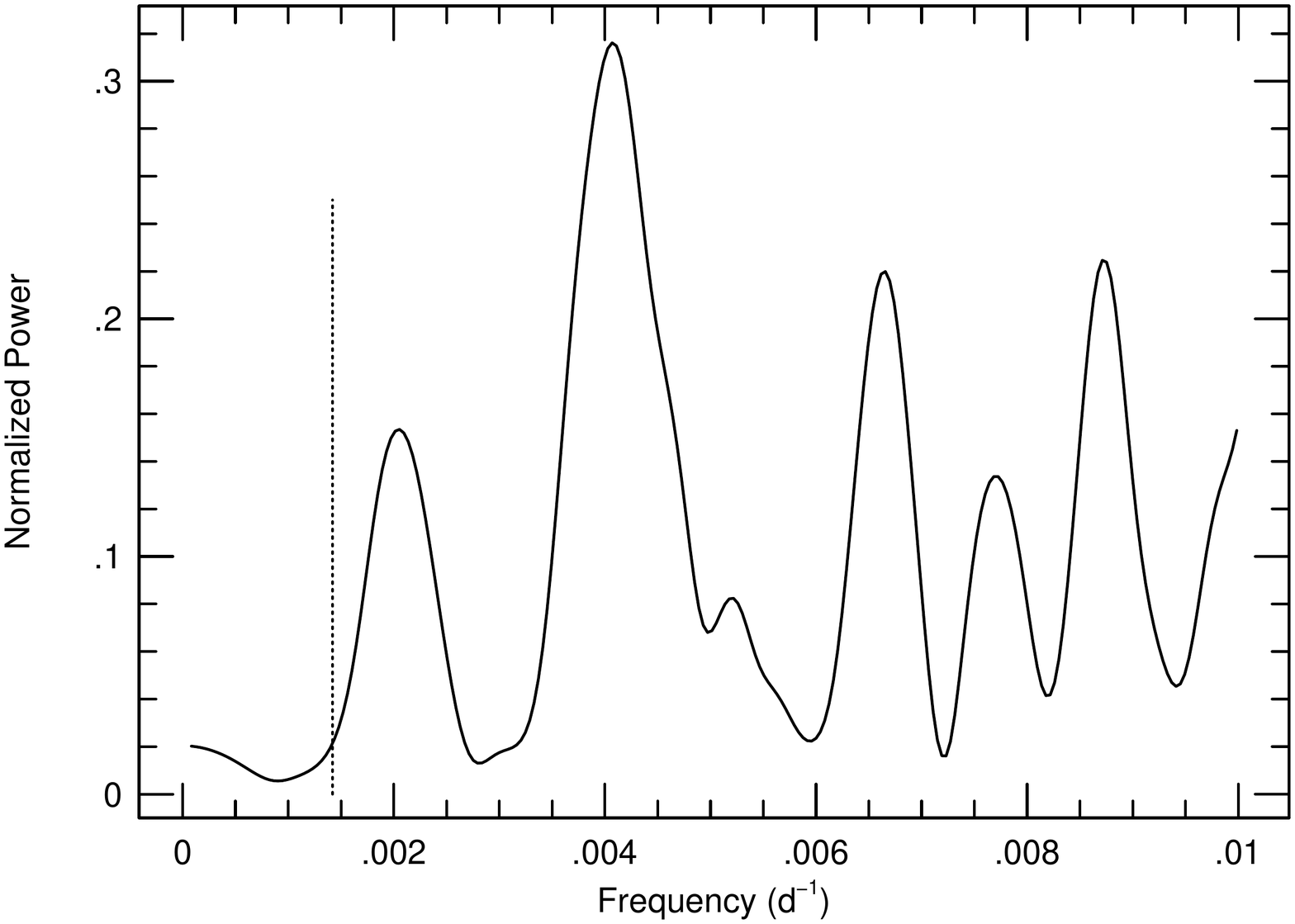}}
\caption{GLS periodogram of the {\rm Hipparcos}  photometry. The vertical line
marks the location of the orbital frequency.
}
\label{ftphot}
\end{figure}

\begin{figure}[h]
\resizebox{\hsize}{!}{\includegraphics{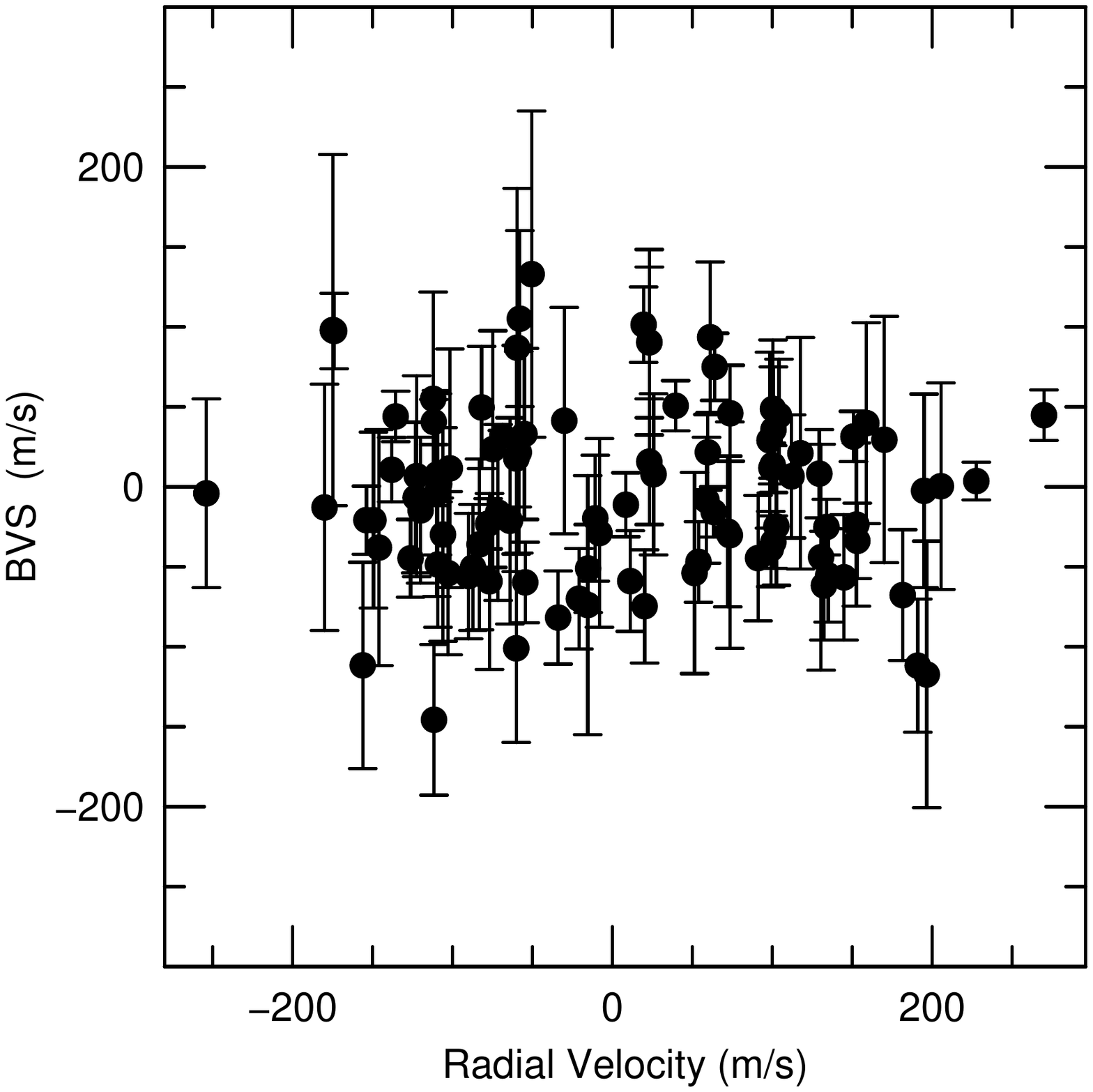}}
\caption{The bisector velocity span (BVS) versus the RV}
\label{bis}
\end{figure}

\begin{figure}[h]
\resizebox{\hsize}{!}{\includegraphics{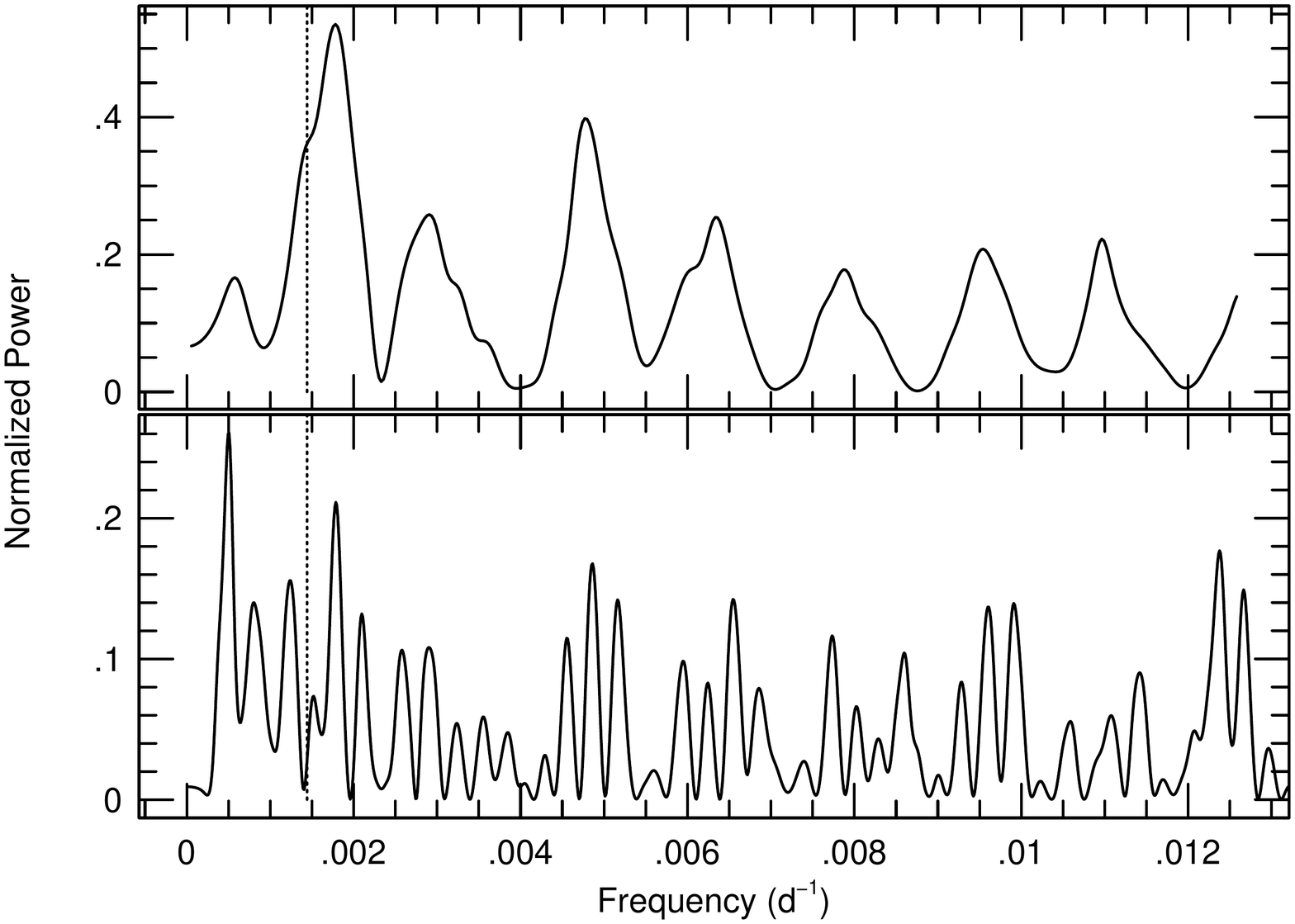}}
\caption{(top) GLS periodogram of the Ca II SMcD-index during 2003 - 2011. 
(bottom) GLS periodogram of the SMcD-index for the full data set. The vertical
dashed line marks the orbital frequency of the purported planet.
}
\label{caft}
\end{figure}

\section{Analysis of RV data}

Figure~\ref{orbit} shows the RV measurements of $\gamma$~Dra from the McD-2.7, BOAO, and TLS
data sets. We will first focus our analysis on data taken from 2003 -- 2011 which show a 
coherent, long-lived periodic signal, but only for the first half of the data.  Afterwards we turn
our analysis to using the full data set.

\begin{figure}[h]
\resizebox{\hsize}{!}{\includegraphics{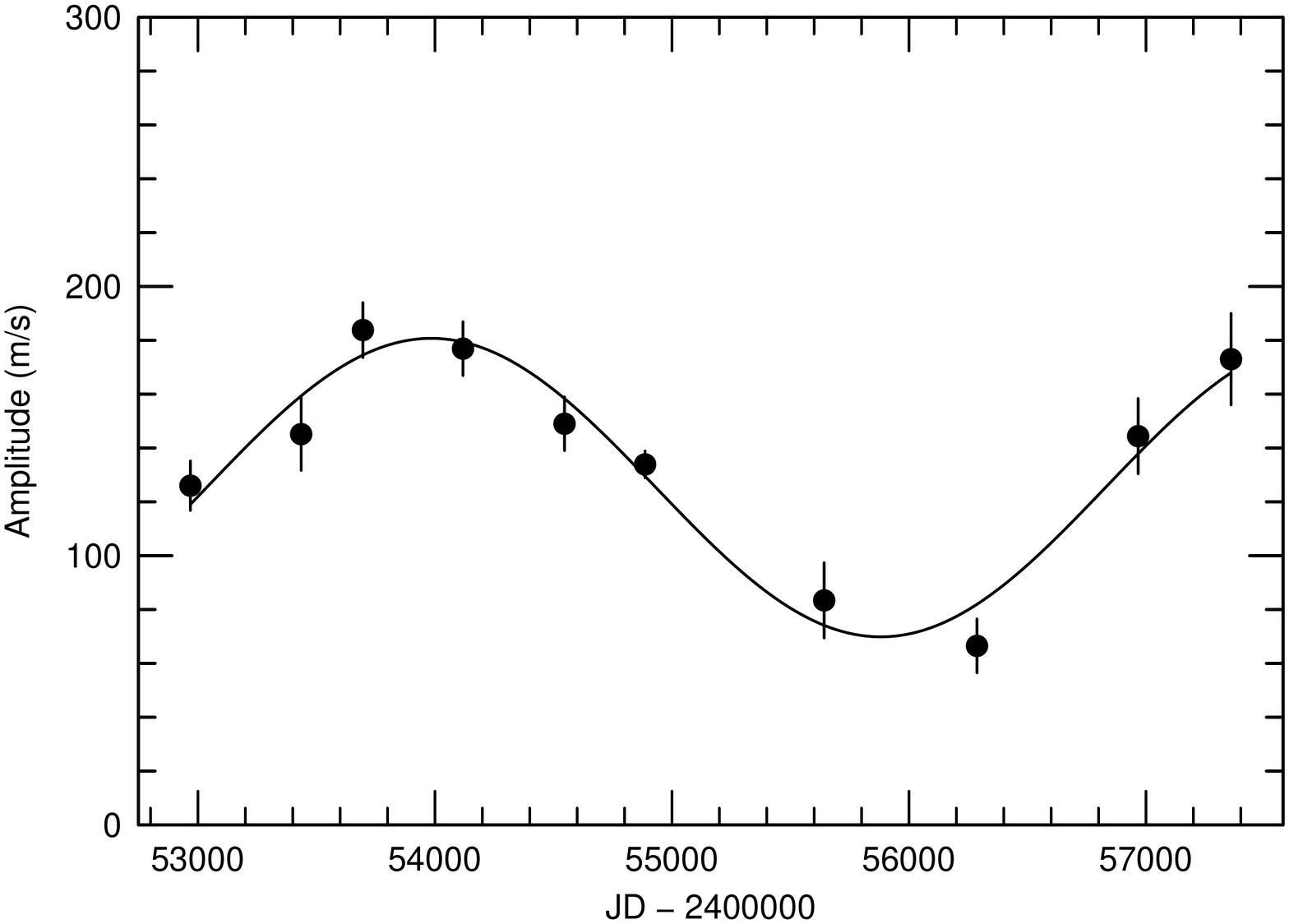}}
\caption{Amplitude variations of the 702-d period found in the RV measurements.
The curve represents a sine-fit with a period of 10.6 years}
\label{ampvar}
\end{figure}

\begin{figure}[h]
\resizebox{\hsize}{!}{\includegraphics{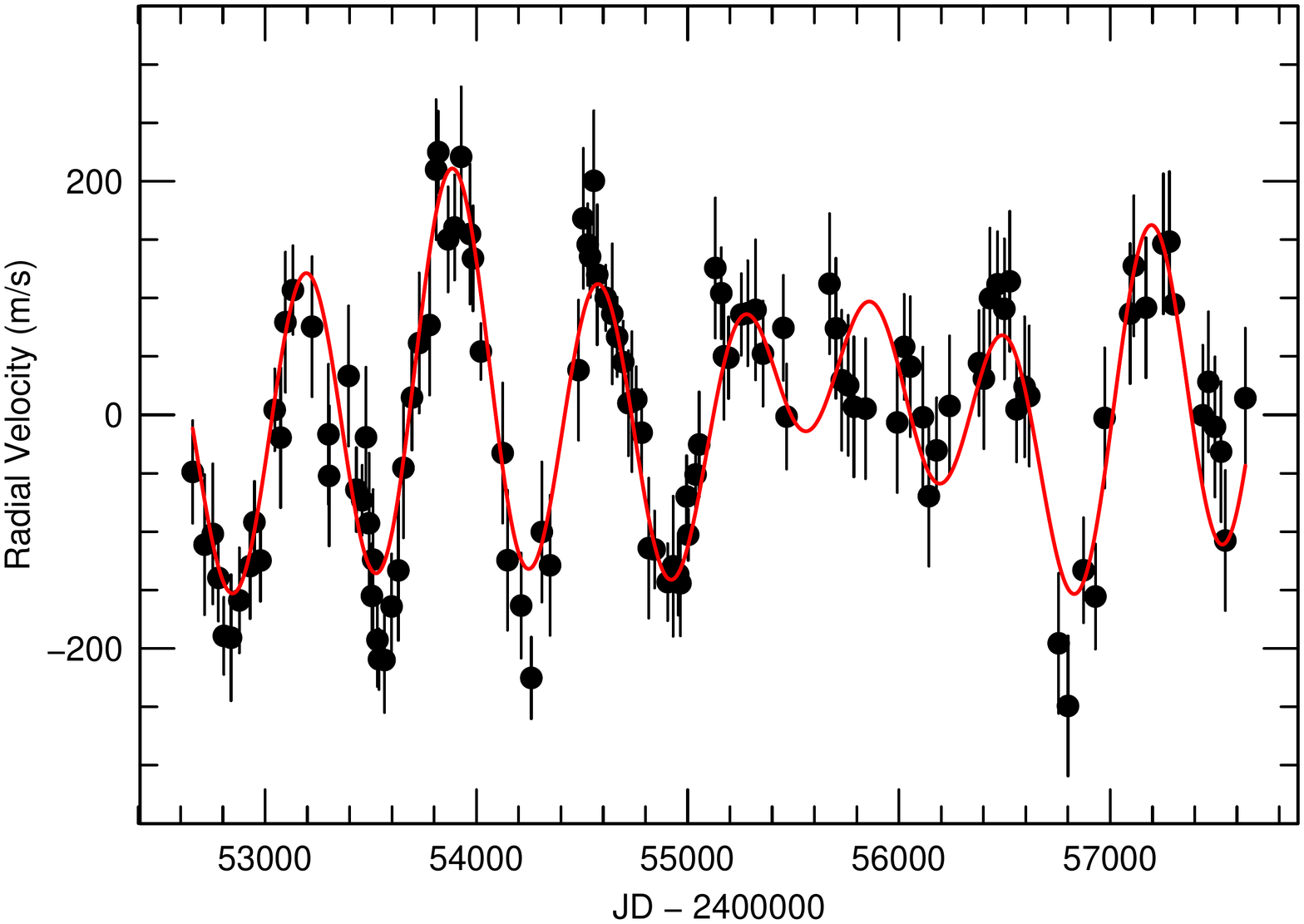}}
\caption{Three component sine-fit (curve) to the RV data from 2003-2017 using the
parameters found in the pre-whitening process (Table 5). The error bars
represent the rms scatter about the solution.}
\label{pulsations}
\end{figure}

\begin{figure}[h]
\resizebox{\hsize}{!}{\includegraphics{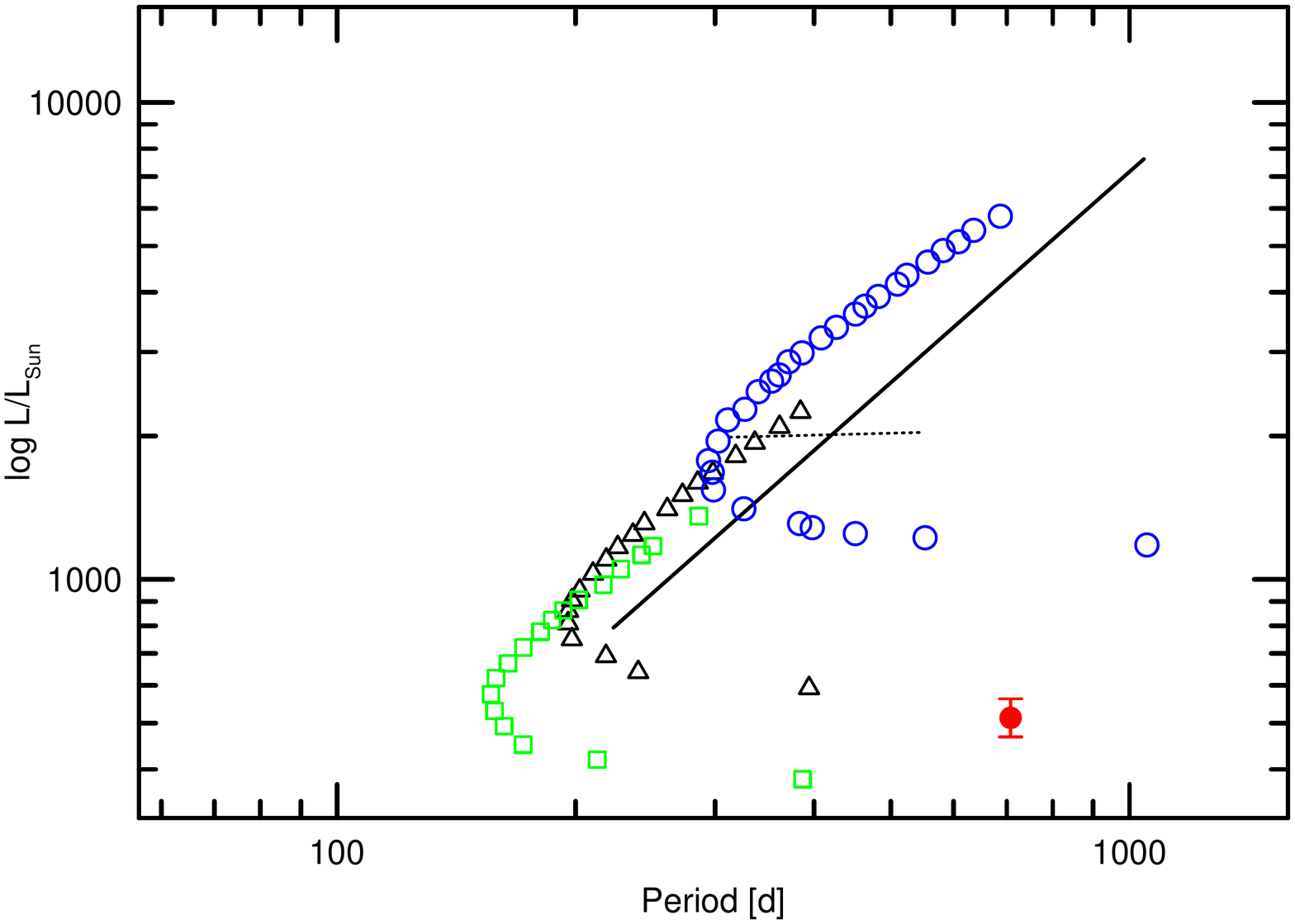}}
\caption{Period-Luminosity relations of dipole oscillatory convection modes (circle) taken from
Saio et al. (2015). The models shown are representative cases calculated
using a mixing length, $\alpha$ = 1.2 and for stellar masses of 1.0 (squares), 1.3 (triangles), and 2.0 (circles)
$M_\odot$. The
diagonal line represents a fit to red giants found in the LMC by Soszyn\'ski et al. (2009). The
horizontal line dashed line represents the approximate spread of variables in the LMC. The dot
marks the location of $\gamma$ Dra.}
\label{period_luminosity}
\end{figure}

\subsection{The 2003 -- 2011 Data: Early evidence of a planetary companion}

A periodogram analysis of the 2003 -- 2013 RV data showed significant
power at a frequency of $\nu$ = 0.00143 d$^{-1}$ (period, $P$ = 699.3 d) which is readily
apparent in the time series.
We fit a Keplerian orbit to these RV data  using the
program {\it Gaussfit} (Jefferys et al. 1988).  
Since each data set has its own zero-point velocity we allowed  this 
to be a free parameter in the fitting process.  Fortunately, there is good temporal overlap
between the three data sets (TLS, McD-2.7, and BOAO).
We note that all  tabulated RV data are with the  individual zero-points subtracted. 

The derived orbital parameters are period, $P$ = 692.4 $\pm$ 2.6 d, radial velocity amplitude,
$K$ = 147.3 1 $\pm$ 4.7 m\,s$^{-1}$, 
eccentricity, $e$ = 0.16 $\pm$ 0.03,
time of periastron, $T_0$ = 2446722.39 $\pm$ 31.06,
and argument of periastron, $\omega$ = 203.7 $\pm$ 13.5 degrees. 

The McD-2.1 RV measurements taken between 1991 and 1993 also show variations
with a period of $\sim$ 700 d (Fig.~\ref{mcdorbit}).
When computing the Scargle periodogram (Scargle 1982) the power, $z$ increases
from $z$ $\approx$ 74 to $z$ $\approx$ 83 after adding the McD-2.1 data. This represents
a decrease in the false alarm probability by a factor of $\approx$ 10$^4$.
The signal was apparently present in the earlier measurements,  so we tried fitting a Keplerian orbit including these
earlier measurements.
This final orbit is shown by the line
in Figs.~\ref{orbit} and ~\ref{mcdorbit}. The parameters are listed in Table~4. The
hypothetical companion has a minimum mass of 10.7 $M_{Jup}$. Such a massive giant planet with a relatively long (few hundreds of days) orbital period is typical 
for planets around K giant stars.

The top panel of Fig.~\ref{phaseorbit} shows the RV variations phased to the orbital period. Even the earlier
McD-2.1 data seem to follow the orbit reasonably well. 
The fourth column in Table 2 shows the rms scatter of the RV data sets about the orbital
solution, but only for the time span 2003 -- 2011.  These have a mean error of $\approx$ 50 m\,s$^{-1}$, significantly larger
than the internal measurement error, but entirely consistent with stellar oscillations.
Using the scaling relationships of Kjeldson \& Bedding (1995) we
estimate a velocity amplitude for the pulsations, $v_{osc}$ $\sim$ 55 m\,s$^{-1}$. The observed  deviations of the RV from the orbital solution can thus be naturally explained by   the intrinsic stellar jitter due to stellar oscillations. 

\subsubsection{``Confirmation'' of the purported planet}

The RV measurements of $\gamma$~Dra taken between 2003 and 2011
show strong evidence of orbital motion due to a planetary companion. However,
as with all such ``planet'' discoveries we  must exclude stellar variability (e.g. rotational modulation, pulsations, etc.) as a cause of the RV behavior. The examinations of 
the spectral line
shapes, photometric variations and activity indicators 
have become common tools for the confirmation of planetary
companions around stars, most notably K giant stars. It is of interest to see if these
tools could have given us an indication that the RV variations of $\gamma$~Dra were due
to stellar variability. For this purpose we will  examine only the time when the star showed clear periodic RV variations.

\subsubsubsection{{\rm Hipparcos}  Photometry}

It is difficult to get precise light curves on very bright stars like $\gamma$~Dra using
ground-based measurements. For discoveries from K giant planet  search programs
one generally has to rely on photometry taken by the {\rm Hipparcos}  space mission.
This photometry is usually not contemporaneous with the RV data, but it is  in the
case  for the early McD-2.1 measurements.
The Generalized Lomb-Scargle (GLS) periodogram (Zechmeister \& K\"urster 2008)
of the {\rm Hipparcos}  photometry shows  no significant power at the orbital frequency (Figure~\ref{ftphot}). The rms scatter of the photometric measurements is only about 4.8 millimag. 

\subsubsubsection{Line Bisector Variations}

Another means of  testing for the planetary nature is to search for variations in the shapes of the
spectral lines with the spectral line bisector (Hatzes, Cochran, \& Johns-Krull 1997; Hatzes, Cochran,
\& Bakker 1998; Queloz et al. 2001). 
Whereas pulsations or stellar surface  structure should produce RV variations that are accompanied by line shape variations,
the orbital motion due to a companion will cause an overall Doppler shift of the spectral line without any changes in the line shape. 

To investigate any changes in the spectral line shapes we first calculated the cross-correlation
function (CCF) using one of our observations as a template, but restricting the calculation 
to the spectral region 4720--4900\,{\AA} which
is largely free of iodine absorption lines.  Only the
TLS data was utilized for several reasons. First, the TLS echelle spectrograph has few moving parts. 
One only has to move different grisms that are mounted on a slide
into the light path in order to access a different wavelength region.
The spectrograph is typically used
in the exact same setup over a time span of several months. This ensures a relatively stable 
instrumental profile (IP) which is important since in measuring the shape of the CCF we are not taking
into account any possible change in the IP. Second, in looking for line shape
variations it is important to have as high a spectral resolution as possible. The TLS data has slightly higher
resolving power over the McD-2.7 data. The BOAO data have  higher spectral resolution, but the IP
 is more unstable compared to the TLS data (Han et al. 2010). The RV calculation takes into account changes in the IP, but 
not so for the bisector measurements.  

We calculated the bisector of the CCF -- the locus of the midpoints calculated from both sides of the
CCF having the same relative flux value. A linear least-squares fit was then made to the
CCF bisectors. We then converted this slope  between the CCF height values of 0.3 and 0.85 to an equivalent velocity which we will call the bisector velocity span (BVS). 
Figure~\ref{bis} shows the correlation of the BVS with the RV 
measurements.  There is no clear correlation between the two quantities as the probability that they
are uncorrelated is 0.48. The 702-d RV variations do not seem to be accompanied by line shape 
variations. Note that there is a large range in the RV variations not seen in the BVS.

\subsubsubsection{Ca II Variations and Rotational Modulation}

The wavelength coverage of the McD-2.7m data includes the Ca II K line. In the past
a ``McDonald S-index" has been used to search for chromospheric variability in stars that are candidates
to host extrasolar planets (e.g. Hatzes et al. 2015). The mean
McDonald  S-index, SMcD-index, should  not to be confused with the
Mt. Wilson S-index (see Paulson et al. 2002 for a definition of the SMcD-index).
Normally, we can use the mean SMcD-index as a measure of the activity level of the star, but this is not the
case for  a giant star such as $\gamma$~Dra due to calibration issues between giants and dwarfs.
However, we are not so much interested in the mean activity level of $\gamma$~Dra, 
but whether 
chromospheric variability might
be related to the RV variations. For our purposes the relative changes in 
the SMcD-index is what is important.

The top panel of Fig.~\ref{caft} shows the GLS of the Ca II SMcD-index during 2003 -- 2011,
the time when the 702-d  variations were present in the RV measurements of 
$\gamma$~Dra. The highest peak occurs at a period of 560 d (frequency = 0.00178 d$^{-1}$),
GLS yields a nominal false alarm probability (FAP) 
of $\approx$ 1 \% for the peak. However, a bootstrap analysis by randomly shuffling the time stamps
of the data and examining the resulting periodograms yielded a much higher
FAP $\approx$ 10 \%. The peak is marginally significant, but if it is a real signal it is clearly displaced
from the orbital frequency. The power of this peak weakens when analyzing the full 
data set, but is still present (lower panel of Fig.~\ref{caft}). If real, we believe that this period
corresponds to the rotational period of the star.

We can  estimate the rotational period, $P_{rot}$,  of the star from the stellar radius and projected rotational
velocity, $v$ sin $i$. Massarotti et al. (2008) measured a  $v$ sin $i$ = 6 km\,s$^{-1}$ for
$\gamma$~Dra. Combined with the interferometric radius of the star yields
$P_{rot}$ $\approx$ 400 d. This  $v$ sin $i$ determination may be somewhat high.
We also estimated the projected rotational velocity using our high-resolution spectra.
We fit a few selected spectral lines with synthetic line profiles that were broadened 
by macroturbulence and rotation. For the macroturbulence we use a value of 6 km\,s$^{-1}$ which
is typical for giant stars with the same surface gravity and temperature of $\gamma$~Dra
(Carney et al. 2008). A good fit to the spectral lines was obtained for 
$v$ sin $i$ $\approx$  4 $\pm$ 1 km\,s$^{-1}$. This yields a rotational period of 600 $\pm$ 150 d.
So, our best guess for the rotational period of $\gamma$~Dra is 400-600 d.
This is consistent with the 560-d period found in the subset SMcD-index measurements. We take this
as the nominal rotational period of the star.

We caution the reader, however, that this estimate of the rotational period is uncertain.
The peak SMcD-index periodogram is not convincingly significant. The estimated rotational  period
also depends on the measured $v$ sin $i$ which is difficult to determine accurately for slowly rotating
stars. Differential rotation, which may be significant in K giant stars, only complicates matters.

The RV data taken up until 2011 show variations long-lived, and coherent for at least 7 years
and arguably 20 years when including the measurements from the 2.1-McD data set. The standard tools of planet confirmation - lack of photometric and line profile
variations - also support the planet hypothesis. Had we stopped our RV measurements in 2011, a logical conclusion would be that the RV variations seen in $\gamma$~Dra arise from the presence of  a signal planetary companion.

It is clear from the lower panel of  Figure~\ref{orbit} that starting in 2011 the 
characteristics of the RV variations abruptly changed. 
Between 2011 and 2014 the RV was relatively constant, or at least there was no evidence for strong periodic variations.
Starting near 2014 the  periodic variations returned with the same
amplitude, but with a clear phase shift of about 0.20. Thus, the RV data from 2011 onward appear
to refute the planetary hypothesis.

\subsection{Amplitude Variations}
The RV measurements of $\gamma$~Dra show  a hint of amplitude variations 
in the 2003-2011 data. For instance, the RV seems to be systematically higher than the
orbital solution during 2006--2007.
To investigate possible amplitude variations we divided the RV variations into slightly overlapping time 
segments.
For each time subset we fit a sine wave to the data keeping the period fixed to 702 d,
but allowing both the amplitude and phase to vary.  Fitting these amplitudes
with a sine function results in a period of $P$ = 3793 $\pm$ 108 d (Fig.~\ref{ampvar}).

\subsection{Frequency analysis of the full data set}

The beating of two closely-spaced
periods can also mimic amplitude variations and these periods should be evident
in a Fourier analysis of the full data set. We performed a frequency analysis 
using the program {\it Period04} (Lenz \& Breger 2005). 
A pre-whitening approach was used, i.e.  the dominant period
in the data was found, a sine-function fit to this and the signal was subtracted. We then searched
the residual RVs and found three significant signals (FAP 
$<$ 0.01) in the RV data  and these are listed in Table 5. To assess the FAP we use the criterion that a peak
having an amplitude at least four times higher than the surrounding noise level has
a FAP $\approx$ 0.01 (Kuschnig et al. 1997). The phase is reckoned starting at 
JD = 2452656.6641. Two of these, $P_1$ = 665.8 d and $P_2$ = 800.9 d
will produce beat-related amplitude variations of 3945 d, very close to the period of the observed
amplitude variations. The fit to the RV data using the three sine components is shown in Figure~\ref{pulsations}.
The rms scatter about this fit is 49 m\,s$^{-1}$. The error bars represent this scatter which is a more
realistic estimate of the RV error that includes intrinsic stellar variability (the so-called ``jitter'').

\section{The Nature of the Multi-periodic Signals}
The observed amplitude variations can be explained by the presence of two signals, $P_1$ and  $P_2$ 
with closely-spaced in period. Two possible hypotheses  for these are
1)  a two-planet system or 2) stellar oscillations.

\subsection{A Two-planet System}

We can denote the semi-major axis
of the outer planet by $a$ = 1 $+$ $\Delta$, where $\Delta$ is the fractional separation, 
and $\mu_1$ and  $\mu_2$ the mass ratios of the two planets with respect to the host star. 
The velocity K-amplitude of $P_1$ and  $P_2$  imply companion masses of $\approx$ 9 M$_{Jup}$
and 3.5 M$_{Jup}$, respectively. 
The orbits
are stable if $\Delta$ $>$ 2.4($\mu_1 + \mu_2$)$^{1/3}$ (Gladman 1993). For $\gamma$~Dra 
2.4($\mu_1 + \mu_2$)$^{1/3}$ = 0.56, 
much larger than the value of $\Delta$ = 0.12. At first glance
the system seems unstable. 

A proper assessment of the stability of the system can only come from numerical simulations.
Possibly there are some configurations for which the system would be stable.
To do this we started from a Keplerian model restricted to circular orbits 
(see Table~5). The best-fit Keplerian solution of a 
two-planet system is very unfavorable for a dynamically stable 
configuration. The rather large planetary masses, the small difference 
in semi-major axes and the absence of a mean motion resonance make it 
unlikely to find long-term stable configurations. In order to estimate 
the fraction of stable configurations, we used the N-body integrator 
{\it Mercury6} (Chambers 1999) to investigate the dynamical stability. 

In a first step, co-planar configurations with $i=90^\circ$ were 
investigated using 1000 random variations of the start parameters 
(mass, mean anomaly, and semi-major axis) within the estimated 
uncertainties. The mass of the central star was kept fixed at 2.14 
M$_{\rm \odot}$. In our calculations, 28 out of the 1000 configurations survived more than 
1Myr, however, most of the configurations were dynamically unstable on 
time scales of few orbits. Given the small difference of the semi-major 
axes this is not surprising. Since not all of the configurations were 
unstable on short time scales, we also allowed for mutual inclinations 
in a second run in order to search in a larger parameter space. Out of 30000 
randomly chosen start configurations 514 survived 1 Myr. This is a lower 
fraction of stable configurations. While the mutual inclinations help to 
avoid close encounters, the increasing mass at lower inclinations of 
the planets increases the gravitational interaction which counterbalances 
the former effect. For four of these we tested the dynamical stability 
over the estimated age of the star (1.6 Gyr). Two out of the four also 
survived. 

Although this is a coarse investigation of the dynamical 
properties of a potential two-planet configuration, we  nevertheless  cannot
outright exclude that a two-planet system is indeed stable. The probability for this 
is low at about  
1-2\%. Given such a  low probability, a more detailed dynamical investigation is 
required, but that is beyond the scope of this paper.

\subsection{Stellar Oscillations?}

The RV variability of $\gamma$ Dra may be related to the Long Period Variables (LPV). These
stars pulsate with periods much longer than the fundamental radial mode. LPVs include,
among others M giant stars, Mira variables as well as semi-periodic variable stars
(SPV). Hinkle et al. (2002) investigated the velocity variability of a sample of nine
LPVs and found that six showed periodic variations with periods, $P$ = 300 -- 900 d and amplitudes
of $\approx$ 2 km\,s$^{-1}$. They proposed that these were a new form of stellar oscillations.

As an example, we take one star from their sample, g Her, that had good stellar
parameters as listed in the {\it Simbad} database. The RV variations had $P$ = 843 d
with a K-amplitude of 2.3 km\,s$^{-1}$. We estimate that g Her
has a luminosity of $L$ = 7300 $L_\odot$ or about 14 times the luminosity
of $\gamma$ Dra. For the sake of argument, let us assume that the ``pulsations'' are related
to high degree p-mode oscillations so that we can apply the scaling relationships
of Kjeldsen \& Bedding (1995). These indicate that the velocity amplitude
of the oscillations, $v_{osc}$, scales as $v_{osc}$ $\propto$ $L$. Scaling from 
the g Her oscillations we thus expect $v_{osc}$ $\sim$ 135 m\,s$^{-1}$ for $\gamma$ Dra, comparable to 
the observed RV amplitude. 

The photometric amplitude can also be estimated from the scaling relationships: 
$\delta L$/$L$ $\propto$ $v_{osc}$. Kiss et al. (1999) measured a photometric amplitude of $\sim$ 0.2
mag for this period. This implies a photometric amplitude of $\sim$ 0.02 for $\gamma$ Dra.
This is comparable to the peak-to-peak variations in the {\rm Hipparcos}  photometry ($\sim$ 0.01 mag),
although this shows no periodic variability. However, given the uncertainties in estimating
photometric variations, it is possible that these have an amplitude lower than the {\rm Hipparcos}  detection limit.

However, it is unlikely that the hypothetical oscillations in $\gamma$ Dra are p-mode oscillations. A more likely
hypothesis is that they may be related to dipole oscillatory convection modes in red giants. Saio et al. (2015) 
proposed these as a possible explanation of the long secondary period in LPVs. Oscillatory
convective modes are non-adiabatic
g$^-$ modes that are present in luminous red giant stars with luminosities log$(L/L_\odot)$ $\gtrapprox$ 3.

The Period-Luminosity (P-L) for radial pulsators lie on roughly a straight line in the log($L$)-log($P$) plane.
In contrast, the P-L relations of oscillatory convective modes have a peculiar shape that bends and becomes
more horizontal for less luminous stars. Figure~\ref{period_luminosity} shows a sample model from
Saio et al. (2015) for stars with three different stellar  masses (1.0, 1.3, and 2.0 $M_\odot$) and a mixing length,
$\alpha$ = 1.2. The diagonal line shows a ``by eye''  fit to the LPV of the so-called ``D'' branch found by
Soszyn\'ski et al. (2009). The location of $\gamma$ Dra seems to be consistent
with the 1.3 $M_\odot$ model if one extrapolates this to lower luminosities.
This model has a higher mass than our inferred mass for $\gamma$ Dra, but given the uncertainties
in this first calculation (convection, metalicity, etc.) the hypothesis that we are seeing
oscillatory convective modes in $\gamma$ Dra seems at least plausible.

\section{Discussion}

The long-term RV monitoring of $\gamma$~Dra shows ostensibly the  ``rise and fall'' of an exoplanet.
The ``rise'' occurs over seven years (2003 -- 2011). This is characterized
by apparently stable, coherent variations with a period of 702 d that are consistent with a planetary companion having a minimum mass of 10.7 $M_{Jup}$. This ``planetary companion'' has characteristics that seem typical for giant planets around K giant stars - massive planets with orbital periods of several hundreds
of days. An examination of the bisector velocity span and photometric measurements from {\rm Hipparcos} 
revealed no variations with the RV period. The case for the planet hypothesis is more credible by the fact 
that the RV variations of this planet seems to have been present in data taken 20 years earlier.
Furthermore, our best estimate of rotational period for the star (560 d) derived
from the Ca II data
suggests, at face value, that the 702-d period is not the rotational period of the star.

The RV measurements taken from 2014--2016, however, contradict this conclusions and thus mark
the ``fall'' of the planet candidate. The periodic RV variations abruptly cease and appear to be absent for the ensuing 3 years. They suddenly appear in 2016 with the same $K$-amplitude and period, but with a clear phase shift of approximately
0.2. It is difficult to reconcile these variations with the planet hypothesis.

There are two explanations for the amplitude variations. One, these may be due to a single period 
whose amplitude is changing with a period of approximately 10.6 yr. This also would have to be accompanied
by phase shift in 2007. A plausible mechanism for this would be the decay of an active region that 
completely disappears by 2011 followed by the emergence of a new one, at a different location
of the star starting in mid-2013. If true, it is puzzling why we do not see any evidence
for this in the standard activity indicators. However, we know very little about  the nature of any
possible surface activity on giant stars and maybe these are not accompanied by measurable
changes in the photometry, line bisectors, or  Ca II emission. 

Alternatively, the amplitude variations can simply arise from the beating of two closely-spaced periods,
$P_1$ = 666 d and $P_2$ = 801 d. This is the simplest and most likely explanation. 
Furthermore, it appears to provide a natural explanation for the phase shift seen in 
2014 (Fig.~\ref{pulsations}). We can only speculate
as to the nature of these periods. 

We cannot completely rule out that these two periods arise from a two-planet system. However, such 
a system would qualify as another ``dynamically challenged" K-giant system. Our simple dynamical
analysis suggests that there is only about 1--2 \% chance that the orbits of the two-planet
system are stable. One could
argue that it is an unlikely system, but we see the system, so nature has found a way to produce it.
Unfortunately, we cannot conclusively refute the 2-planet hypothesis.

On the other hand, one could argue that either $P_1$ or $P_2$ is due to a planetary companion,
and that the other is due to rotational modulation of surface features. The 666-d period can be
consistent with the generous estimate of the rotational period  of 400-600 d.
Again, the rotational modulation in the RV  would have to produce no variations
in other quantities (bisector, Ca II, photometry).
Although we cannot exclude this hypothesis, we do not favor it for
the simple reason that whenever a beat phenomena between two closely
oscillations is a prime suspect. At the present time
we favor stellar oscillations  as the cause of $P_1$ and $P_2$.

The frequency analysis of the full data set reveals yet a third period, $P_3$, at 1855 d which is
significant. We cannot exclude that this signal is in fact due to a planetary companion with a minimum
mass of 2.8 $M_{Jup}$. However, given the fact that we have found two other long periods that are most likely
due to stellar oscillations, it would be premature at this time to attribute a third long period to a
companion. Continued monitoring of the star should be able to verify this.

Hatzes et al. (2015) reported two periods (629 d and 520 d) in the RV measurements of the K giant
star $\alpha$~Tau that spanned 30 years. The longer period was attributed to the presence of a planetary
companion while the 520-d period was interpreted as rotational modulation. Given that the periods
found in $\alpha$~Tau are comparable to those in $\gamma$~Dra and both stars are evolved with large
radii, a closer scrutiny of the RV variability of $\alpha$~Tau is warranted.

The RV variations that we have discovered in $\gamma$~Dra are both troubling and exciting. The troubling
aspect stems from the fact that seven years of RV measurements for this star showed clear periodic
variations that were long-lived and coherent. An orbital fit yielded a period and companion
mass that was typical for planets around K giant stars. The scatter about the orbital
fit could easily be explained by short-term oscillations on time scales of days and night-to-night
variations of 50 m\,s$^{-1}$ or more (see Table 3). An examination of the {\rm Hipparcos} photometry and line
bisectors showed no variations with the planet period. 
The standard tools used to confirm planet discoveries seem to have failed in this case.

One can only speculate as to how many of the giant planets around K giant stars may actually be due
to stellar variability like the one we have found for $\gamma$~Dra. Can this new phenomenon
explain some of the ``problem'' multi-planets around K giant stars? Are there other instances
where this new-found stellar variability is masquerading as a planet? In the case
of $\gamma$~Dra using seven years of observations we would have arrived at a logical conclusion (planetary
companion) that may be wrong. Only after long-term monitoring ($\sim$ decade) 
did the star reveal a more complicated nature for these
variations.

The exciting aspect of our RV measurements is that we may have stumbled upon a new phenomenon
in K giant stars, and quite possibly a new type of stellar oscillations. More studies are needed
to investigate the nature of these RV variations. In particular, how these correlate with the
evolutionary status of the star. Interestingly, $\gamma$~Dra is highly evolved and with a large radius.
Perhaps this holds the key to understanding these variations. When looking for exoplanets
around giant stars with the RV method, long-term monitoring over many years is required. 
Our observations of $\gamma$ Dra only highlights that the community should be more critical in examining
the nature of long-period RV variations in K giants and to not be so eager to attribute these to 
exoplanets.

\begin{acknowledgements}
This research has made use of the SIMBAD data base operated
at CDS, Strasbourg, France. APH acknowledges the support of DFG grants HA 3279/5-1 and HA 3279/8-1.
The McDonald Observatory Planet Search (ME, WDC, PJM) is supported by the National 
Science Foundation through grant AST-1313075.

\end{acknowledgements}

\begin{table}
\begin{center}
\begin{tabular}{ll}
\hline
\hline
Parameter  & Value \\
\hline
$T_{\rm{eff}}$  [K]             &   3990  $\pm$ 42 K$^{1}$\\
 log $g$                              & 1.67 $\pm$ 0.1$^{1}$  \\
 $[$Fe/H$]$                        & $+$0.11 $\pm$ 0.05$^{1}$  \\
 $v$sin$i$ [km\,s$^{-1}$]  &  4.5 $\pm$ 0.05 \\
 Mass  [$M_\odot$]           & 2.14 $\pm$ 0.16  $M_\odot$ \\
 Angular Diameter [mas]  & 9.86 $\pm$ 0.128 \\
Parallax [mas]                    & 21.14 $\pm$ 0.10\\
 Radius  [$R_\odot$]         & 49.03 $\pm$ 2.52 $R_\odot$ \\
 Age			                  & 1.30 $\pm$ 0.25 Gyr \\
$V$-mag                             & 2.23 $\pm$ 0.009 \\
$B-V$                                  & 1.52 \\
$L$ [$L_\odot$]                 &  510 $\pm$ 51 \\
\hline
\end{tabular}
\caption{Stellar Parameters for $\gamma$~Dra  ($^1$Koleva \& Vazdekis  (2012),
$^2$Mozurkewich et al. (2003)).}
\end{center}
\label{star}
\end{table}

\begin{table}
\begin{center}
\begin{tabular}{ccccc}
\hline
\hline
Data Set  &  Coverage & N & $\sigma_{RV}$   \\
          &  (Years)  &   & (m\,s$^{-1}$) \\
\hline
McD-2.1    & 1991.45--1993.58 & 35      & 64.1  \\
McD-2.7    & 2005.26--2016.69 & 65      & 41.3  \\
BOAO             & 2003.38--2015.76 & 82      & 57.5 \\
TLS              & 2003.04--2013.57 & 136      & 47.1\\
\hline
\end{tabular}
\caption{The data sets used in the orbital solution.}
\end{center}
\label{data}
\end{table}

\begin{table}
\begin{center}
\begin{tabular}{lrrc}
Julian Day   & RV (m\,s$^{-1}$)  & $\sigma$ (m\,s$^{-1}$) & Dataset  \\
\hline
2448422.8047  & -30.13 &    10.6 & McD-2.1  \\
2448423.7305  & -53.03 &    6.6  & McD-2.1 \\
2448424.7188  & -114.12&    7.0  & McD-2.1 \\
2448469.8125  & -156.97&    5.5  & McD-2.1 \\
2448523.6406  & -270.98&    12.4 & McD-2.1 \\
2448558.5547  & -227.34&    9.9  & McD-2.1 \\
2448637.0312  & -88.92 &    19.8 & McD-2.1 \\
2448702.0156  & 3.13   &    28.1 & McD-2.1 \\
2448703.9336  & 68.52  &    19.8 & McD-2.1 \\
2448704.8594  & 8.29   &    14.6 & McD-2.1  \\
2448722.9062  & -11.24 &    14.0 & McD-2.1 \\
2448723.9492  & -59.35 &    21.2 & McD-2.1 \\
2448724.9023  & 9.06   &    10.7 & McD-2.1 \\
2448759.9336  & -32.54 &    4.7  & McD-2.1 \\
2448768.8125  & 29.69  &    15.0 & McD-2.1 \\
2448785.8242  & -2.20  &    10.2 & McD-2.1 \\
2448786.8398  & 51.03  &    8.9  & McD-2.1 \\
2448787.7891  & 108.23 &    5.5  & McD-2.1 \\
2448788.7539  & 96.43  &    7.8  & McD-2.1 \\
2448789.7852  & 24.33  &    4.4  & McD-2.1 \\
2448790.7773  & 6.89   &    7.2  & McD-2.1 \\
2448791.7656  & 70.90  &    5.7  & McD-2.1 \\
2448792.7188  & 18.73  &    6.3  & McD-2.1 \\
2448878.5820  & -14.54 &    6.9  & McD-2.1 \\
2448936.5859  & 96.78  &   17.7  & McD-2.1 \\
2449020.0312  & 134.83 &   15.0  & McD-2.1 \\
2449097.9180  & 109.35 &    6.2  & McD-2.1 \\
2449098.9375  & 48.95  &   7.5   & McD-2.1 \\
2449099.8906  & 61.57  &   4.4   & McD-2.1 \\
2449118.8867  & 44.23  &   8.2   & McD-2.1 \\
2449119.7773  & -0.99  &   6.7   & McD-2.1 \\
2449198.8008  & -31.88 &   5.3   & McD-2.1 \\
\hline \\
\end{tabular}
\caption{The RV Data}
\end{center}
\label{McD-2.1}
\end{table}

\begin{table}
\setcounter{table}{2}
\begin{center}
\begin{tabular}{lrrc}
Julian Day   & RV (m\,s$^{-1}$)  & $\sigma$ (m\,s$^{-1}$) & Dataset \\
\hline
2449199.8086  & -91.47 &  12.4 & McD-2.1   \\
2449200.7930  & -93.08 &  10.5 & McD-2.1  \\
2449201.7617  & -56.20 &  9.8  & McD-2.1 \\
\hline
2453463.9492   & -82.24  & 4.23 & McD-2.7m \\
2453503.9023   & -185.06 & 6.03 & McD-2.7m \\
2453529.9492   & -226.13 & 2.83 & McD-2.7m \\
2453530.9296   & -207.21 & 2.81 & McD-2.7m \\
2453531.7968   & -172.82 & 2.86 & McD-2.7m \\
2453532.8593   & -124.47 & 3.61 & McD-2.7m \\
2453533.8476   & -225.33 & 2.86 & McD-2.7m \\
2453534.8320   & -230.59 & 2.62 & McD-2.7m \\
2453535.8320   & -182.47 & 2.23 & McD-2.7m \\
2453536.8320   & -191.81 & 3.41 & McD-2.7m \\
2453537.8593   & -227.78 & 3.53 & McD-2.7m \\
2453563.7695   & -203.6  & 5.06 & McD-2.7m \\
2453630.7031   & -133.1  & 4.42 & McD-2.7m \\
2453654.6640   & -45.35  & 5.97 & McD-2.7m \\
2453689.5234   & 1.60    & 4.12 & McD-2.7m \\
2453864.9257   & 121.91  & 5.82 & McD-2.7m \\
2453907.8203   & 185.28  & 5.99 & McD-2.7m \\
2453927.7812   & 220.86  & 5.65 & McD-2.7m \\
2453969.7539   & 154.81  & 6.40 & McD-2.7m \\
2454019.5507   & 21.55   & 6.47 & McD-2.7m \\
2454020.6406   & 53.36   & 6.11 & McD-2.7m \\
2454309.6367   & -100.30 & 3.83 & McD-2.7m \\
2454348.6250   & -128.82 & 3.32 & McD-2.7m \\
2454555.9492   & 200.35  & 3.93 & McD-2.7m \\
2454571.9101   & 119.94  & 3.73 & McD-2.7m \\
2454604.8593   & 107.64  & 8.21 & McD-2.7m \\
2454606.8632   & 128.94  & 8.96 & McD-2.7m \\
2454665.7304   & 104.19  & 7.08 & McD-2.7m \\
2454965.9492   & -95.17  & 4.63 & McD-2.7m \\
\hline 
\end{tabular}
\caption{The RV Data}
\end{center}
\label{McD-2.7mcd}
\end{table}

\begin{table}
\setcounter{table}{2}
\begin{center}
\begin{tabular}{lrrc}
Julian Day   & RV (m\,s$^{-1}$)  & $\sigma$ (m\,s$^{-1}$) & Dataset \\
\hline
2454990.8437   & -87.09  & 4.25 & McD-2.7m \\
2455288.0000   & 115.67  & 6.62 & McD-2.7m \\
2455701.8750 &  74.10  & 3.26 & McD-2.7m \\
2455759.8007 &  25.25  & 3.41 & McD-2.7m \\
2455786.7968 &  6.87   & 5.06 & McD-2.7m \\
2455992.0234 &  -6.51  & 5.12 & McD-2.7m \\
2456024.8945 &  90.47  & 7.15 & McD-2.7m \\
2456053.9687 &  41.26  & 4.61 & McD-2.7m \\
2456113.6953 &  -2.05  & 6.19 & McD-2.7m \\
2456141.7070 &  -69.63 & 3.62 & McD-2.7m \\
2456179.6835 &  -33.33 & 6.62 & McD-2.7m \\
2456239.5429 &  7.63   & 4.53 & McD-2.7m \\
2456379.8867 &  43.32  & 3.54 & McD-2.7m \\
2456401.9726 &  31     & 3.20 & McD-2.7m \\
2456430.8828 &  99.88  & 4.06 & McD-2.7m \\
2456465.7968  & 112.98 & 3.46 & McD-2.7m \\
2456467.7304 &  110.62 & 3.91 & McD-2.7m \\
2456498.8867 &  90.70  & 4.12 & McD-2.7m \\
2456524.6054 &  114.32 & 4.38 & McD-2.7m \\
2456561.5625 &  -14.10 & 6.30 & McD-2.7m \\
2456594.6015 &  24.02  & 4.54 & McD-2.7m \\
2456755.9882 &  -195.58& 3.67 & McD-2.7m \\
2456861.7343 &  -147.14& 4.65 & McD-2.7m \\
2456885.6093 &  -119.03& 4.67 & McD-2.7m \\
2456938.6562 &  -94.063& 3.71 & McD-2.7m \\
2456973.5312 &  -2.6401& 4.24 & McD-2.7m \\
2457111.9882 &  127.59 & 5.00 & McD-2.7m \\
2457251.6250 &  146.42 & 3.60 & McD-2.7m \\
2457279.6992 &  151.27 & 4.86 & McD-2.7m \\
2457280.5781 &  145.24 & 4.62 & McD-2.7m \\
2457437.9687 &  -0.36  & 4.38 & McD-2.7m \\
2457465.0156 &  28.28  & 4.15 & McD-2.7m \\
2457495.9804 &  -10.35 & 3.40 & McD-2.7m \\
2457524.8515 &  -31.51 & 4.22 & McD-2.7m \\
2457544.8632 &  -107.57& 4.74 & McD-2.7m \\
2457639.6445 &  14.3   & 4.67 & McD-2.7m \\
\hline 
\end{tabular}
\caption{The RV Data}
\end{center}
\label{McD-2.7mcd}
\end{table}

\begin{table}
\setcounter{table}{2}
\begin{center}
\begin{tabular}{lrrc}
Julian Day   & RV (m\,s$^{-1}$)  & $\sigma$ (m\,s$^{-1}$) & Dataset \\
\hline
2452776.0898 & -79.01  & 5.62 & BOAO \\
2452777.1054 & -139.63 & 4.45 & BOAO \\
2452778.2460 & -125.98 & 4.16 & BOAO \\
2452779.1093 & -119.86 & 6.75 & BOAO \\
2452780.2148 & -138.01 & 3.27 & BOAO \\
2452781.1367 & -173.94 & 4.94 & BOAO \\
2452948.9335 & -109.11 & 4.77 & BOAO \\
2452949.9375 & -89.51 & 6.68 & BOAO \\
2452975.9257 & -133.97 & 5.08 & BOAO \\
2452977.9023 & -120.01 & 4.42 & BOAO \\
2452980.8867 & -117.38 & 6.78 & BOAO \\
2453044.3593 & -22.86 & 5.09 & BOAO \\
2453045.3281 & 42.25 & 3.67 & BOAO \\
2453046.3203 & -3.81 & 4.27 & BOAO \\
2453073.3359 & -18.61 & 5.05 & BOAO \\
2453096.2304 & 80.22 & 3.96 & BOAO \\
2453130.3046 & 86.90 & 2.09 & BOAO \\
2453131.2265 & 157.84 & 3.28 & BOAO \\
2453132.2617 & 116.42 & 3.79 & BOAO \\
2453133.0898 & 69.59 & 4.52 & BOAO \\
2453299.9687 & -15.76 & 8.02 & BOAO \\
2453302.9531 & -51.36 & 4.87 & BOAO \\
2453395.3437 & 34.19 & 4.09 & BOAO \\
2453430.3593 & -0.31 & 4.13 & BOAO \\
2453433.3554 & -23.97 & 3.48 & BOAO \\
2453459.1640 & -40.81 & 5.90 & BOAO \\
2453460.2734 & -69.26 & 4.13 & BOAO \\
2453507.2187 & -124.16 & 4.38 & BOAO \\
2453545.1132 & -233.91 & 6.10 & BOAO \\
2453700.9531 & 29.04 & 6.64 & BOAO \\
2453728.8906 & 62.39 & 5.70 & BOAO \\
2453778.3398 & 77.79 & 4.56 & BOAO \\
\hline
\end{tabular}
\caption{The RV Data}
\end{center}
\label{boao}
\end{table}

\begin{table}
\setcounter{table}{2}
\begin{center}
\begin{tabular}{lrrc}
Julian Day   & RV (m\,s$^{-1}$)  & $\sigma$ (m\,s$^{-1}$) & Dataset\\
\hline
2453809.3671 & 210.79 & 4.45 & BOAO \\
2453818.3203 & 249.19 & 4.17 & BOAO \\
2453819.3046 & 218.99 & 6.20 & BOAO \\
2453821.2578 & 209.49 & 5.80 & BOAO \\
2453867.1757 & 179.32 & 3.57 & BOAO \\
2453888.0546 & 109.29 & 7.30 & BOAO \\
2453891.1562 & 112.49 & 7.40 & BOAO \\
2453896.2148 & 188.69 & 6.60 & BOAO \\
2453899.2382 & 209.79 & 6.10 & BOAO \\
2453981.0820 & 152.19 & 6.40 & BOAO \\
2454017.9570 & 75.04 & 4.525& BOAO \\
2454026.9531 & 68.59 & 5.400 & BOAO \\
2454124.3828 & -31.81 & 6.80 & BOAO \\
2454147.3750 & -123.56 & 4.36 & BOAO \\
2454210.2617 & -214.61 & 7.20 & BOAO \\
2454213.3125 & -110.21 & 7.50 & BOAO \\
2454262.2656 & -229.11 & 5.05 & BOAO \\
2454264.2734 & -224.91 & 5.02 & BOAO \\
2454483.4179 & 39.165 & 3.01 & BOAO \\
2454506.3750 & 169.11 & 3.15 & BOAO \\
2454536.3242 & 144.01 & 3.31 & BOAO \\
2454619.0820 & 117.94 & 5.55 & BOAO \\
2454643.2539 & 87.34 & 5.19 & BOAO \\
2454719.0000 & 7.99 & 4.73 & BOAO \\
2454720.0703 & 13.74 & 3.95 & BOAO \\
2454735.9843 & 12.19 & 4.70 & BOAO \\
2454752.0234 & 34.19 & 4.49 & BOAO \\
2454756.0312 & 49.49 & 6.40 & BOAO \\
2454847.3906 & -99.71 & 4.27 & BOAO \\
2454931.1484 & -128.71 & 4.34 & BOAO \\
2454971.2460 & -191.61 & 6.70 & BOAO \\
2454995.2109 & -69.71 & 5.90 & BOAO \\
2455130.9492 & 126.62 & 3.31 & BOAO \\
2455248.3828 & 95.29 & 5.70 & BOAO \\
\hline
\end{tabular}
\caption{The RV Data}
\end{center}
\label{boao}
\end{table}

\begin{table}
\setcounter{table}{2}
\begin{center}
\begin{tabular}{lrrc}
Julian Day   & RV (m\,s$^{-1}$)  & $\sigma$ (m\,s$^{-1}$) & Dataset\\
\hline
2455321.3164 & 90.79   & 5.50 & BOAO \\
2455356.1679 & 74.19   & 6.00 & BOAO \\
2455455.9765 & 127.89  & 6.0 & BOAO \\
2455671.2773 & 113.19  & 6.10 & BOAO \\
2455729.2617 & 30.49   & 5.80 & BOAO \\
2455841.9804 & 6.19    & 4.06 & BOAO \\
2456024.2500 & 26.89   & 7.20 & BOAO \\
2456176.9765 & -26.61  & 4.27 & BOAO \\
2456378.1875 & 46.19   & 4.20 & BOAO \\
2456551.9609 & 24.39   & 6.00 & BOAO \\
2456616.8945 & 16.99   & 6.10 & BOAO \\
2456800.1367 & -248.31 & 6.80 & BOAO \\
2456920.9765 & -216.01 & 7.80 & BOAO \\
2457094.2304 & 87.59   & 7.70 & BOAO \\
2457169.2578 & 92.59   & 8.20 & BOAO \\
2457300.9296 & 95.49   & 5.80 & BOAO \\
\hline
2452656.6640 & -51.53  & 5.95 & TLS  \\
2452657.6015 & -24.90  & 6.28 & TLS  \\
2452713.5937 & -102.04 & 6.93 & TLS  \\
2452752.4609 & -88.57  & 5.59 & TLS  \\
2452802.3828 & -144.46 & 4.17 & TLS  \\
2452804.3671 & -180.28 & 5.91 & TLS  \\
2452807.3671 & -211.63 & 4.01 & TLS  \\
2452835.3867 & -181.19 & 3.46 & TLS  \\
2452836.3750 & -285.53 & 3.25 & TLS  \\
2452838.4687 & -142.71 & 3.69 & TLS  \\
2452839.4218 & -141.16 & 3.47 & TLS  \\
2452840.4375 & -158.05 & 3.53 & TLS  \\
2452811.3593 & -177.85 & 2.85 & TLS  \\
2452877.3125 & -115.81 & 3.58 & TLS  \\
2452878.3203 & -182.02 & 4.03 & TLS \\
2452928.2500 & -132.24 & 3.73 & TLS \\
\hline
\end{tabular}
\caption{The RV Data}
\end{center}
\label{boao}
\end{table}

\begin{table}
\begin{center}
\setcounter{table}{2}
\begin{tabular}{lrrc}
Julian Day   & RV (m\,s$^{-1}$)  & $\sigma$ (m\,s$^{-1}$) & Dataset\\
\hline
2452929.2500 & -104.39 & 3.65  & TLS \\
2452949.2421 & -65.93  & 3.74  & TLS \\
2453222.3867 & 87.45   & 2.56  & TLS \\
2453432.5429 & -155.44 & 3.13  & TLS \\
2453454.6796 & -86.94  & 3.99  & TLS \\
2453477.5625 & -5.75   & 2.76  & TLS \\
2453491.5078 & -84.56  & 6.40  & TLS \\
2453511.4140 & -112.12 & 3.11  & TLS \\
2453528.4179 & -150.35 & 3.79  & TLS \\
2453566.3554 & -202.68 & 1.61  & TLS  \\
2453598.3750 & -141.15 & 2.63  & TLS \\
2453599.3359 & -167.32 & 2.86  & TLS \\
2453873.5078 & 242.19  & 4.01  & TLS \\
2453877.4296 & 173.42  & 2.95  & TLS \\
2453986.4375 & 127.05  & 2.66  & TLS \\
2454253.3750 & -211.41 & 4.31  & TLS \\
2454520.5859 & 165.29  & 12.95 & TLS \\
2454529.7031 & 155.51  & 11.35 & TLS \\
2454530.5078 & 148.95  & 15.28 & TLS \\
2454538.6015 & 162.47  & 11.78 & TLS \\
2454539.5117 & 121.29  & 14.63 & TLS \\
2454614.5742 & 65.91   & 11.79 & TLS \\
2454615.5468 & 100.80  & 9.50  & TLS \\
2454662.3750 & 93.39   & 2.35  & TLS \\
2454663.5781 & 61.90   & 2.58  & TLS \\
2454667.5546 & 22.02   & 2.48  & TLS \\
2454671.3671 & 74.42   & 2.62  & TLS \\
2454694.3906 & 65.87   & 2.64  & TLS \\
2454695.4140 & 62.73   & 1.94  & TLS \\
2454696.3359 & 22.82   & 2.64  & TLS \\
2454759.2343 & 16.97   & 2.12  & TLS \\
2454760.2265 & -13.25  & 2.63  & TLS \\
\hline
\end{tabular}
\caption{The RV Data}
\end{center}
\label{tls}
\end{table}

\begin{table}
\setcounter{table}{2}
\begin{center}
\begin{tabular}{lrrc}
Julian Day   & RV (m\,s$^{-1}$)  & $\sigma$ (m\,s$^{-1}$) & Data Set\\
\hline
2454781.2421 & -45.85  & 3.04 & TLS \\
2454782.3046 & -12.66  & 2.53 & TLS \\
2454783.2304 &  28.75  & 2.49 & TLS \\
2454815.7421 & -108.13 & 3.27 & TLS \\
2454840.7500 & -145.79 & 3.66 & TLS \\
2454841.7539 & -95.51  & 3.05 & TLS \\
245484McD-2.7578 & -65.23  & 3.14 & TLS \\
2454843.7539 & -148.86 & 3.30 & TLS \\
2454845.7578 & -92.58  & 3.21 & TLS \\
2454902.6953 & -97.77  & 3.13 & TLS \\
2454904.6601 & -146.65 & 4.28 & TLS \\
2454908.6796 & -158.51 & 3.09 & TLS \\
2454952.3671 & -124.95 & 4.30 & TLS \\
2454954.3281 & -123.82 & 2.97 & TLS \\
2454959.3203 & -141.60 & 3.19 & TLS \\
2454960.5625 & -138.76 & 3.31 & TLS \\
2454999.3906 & -46.32  & 3.02 & TLS \\
2455000.4648 & -59.69  & 4.63 & TLS \\
2455001.4140 & -113.96 & 2.78 & TLS \\
2455002.4140 & -113.87 & 2.97 & TLS \\
2455003.5156 & -100.24 & 2.90 & TLS \\
2455004.3906 & -96.63  & 3.15 & TLS \\
2455035.3750 & -91.25  & 2.79 & TLS \\
2455037.3750 & -17.32  & 2.64 & TLS \\
2455038.3750 & -26.16  & 2.57 & TLS \\
2455039.4648 & -48.45  & 2.13 & TLS \\
2455051.3437 & 14.03   & 2.83 & TLS \\
2454267.4453 & -30.12  & 1.89 & TLS \\
2454253.3750 & -211.49 & 4.18 & TLS \\
2454840.7500 & -146.43 & 3.74 & TLS \\
2454841.7539 & -95.15  & 3.04 & TLS \\
245484McD-2.7578 & -65.59  & 3.14 & TLS \\
\hline
\end{tabular}
\caption{The RV data}
\end{center}
\label{tls}
\end{table}

\begin{table}
\setcounter{table}{2}
\begin{center}
\begin{tabular}{lrrc}
Julian Day   & RV (m\,s$^{-1}$)  & $\sigma$ (m\,s$^{-1}$) \\
\hline
2454843.7539 & -148.69 & 3.23 & TLS \\
2454845.7578 & -92.35  & 3.18 & TLS \\
2454902.6953 & -97.81  & 3.09 & TLS \\
2454904.6601 & -146.57 & 4.28 & TLS \\
2454908.6796 & -158.51 & 3.09 & TLS \\
2454952.3671 & -123.87 & 4.38 & TLS \\
2454954.3281 & -123.96 & 2.92 & TLS \\
2454959.3203 & -140.77 & 3.16 & TLS \\
2454960.5625 & -139.74 & 3.41 & TLS \\
2454999.3906 & -46.12  & 3.00 & TLS \\
2455000.4648 & -59.28  & 4.65 & TLS \\
2455001.4140 & -113.97 & 2.78 & TLS \\
2455002.4140 & -113.87 & 2.97 & TLS \\
2455003.5156 & -100.24 & 2.90 & TLS \\
2455004.3906 & -96.57  & 3.14 & TLS \\
2455035.3750 & -91.79  & 2.77 & TLS \\
2455037.3750 & -16.83  & 2.65 & TLS \\
2455038.3750 & -26.15  & 2.63 & TLS \\
2455039.4648 & -48.32  & 2.11 & TLS \\
2455051.3437 & 14.60   & 2.84 & TLS \\
2455057.3320 & -54.30  & 2.54 & TLS \\
2455155.3046 & 198.13  & 2.31 & TLS \\
2455157.3398 & 106.63  & 6.52 & TLS \\
2455158.2265 & 100.18  & 2.29 & TLS \\
2455161.2265 & 125.29  & 2.06 & TLS \\
2455162.2734 & 104.79  & 2.36 & TLS \\
2455163.2343 & 71.02   & 2.13 & TLS \\
2455168.2500 & 9.95    & 2.77 & TLS \\
2455170.2812 & 139.16  & 2.71 & TLS \\
2455173.2460 & 61.14   & 2.68 & TLS \\
2455175.2500 & 45.02   & 2.69 & TLS \\
2455192.1796 & 40.14   & 2.99 & TLS \\
\hline
\end{tabular}
\caption{The RV data}
\end{center}
\label{tls}
\end{table}

\begin{table}
\setcounter{table}{2}
\begin{center}
\begin{tabular}{lrrc}
Julian Day   & RV (m\,s$^{-1}$)  & $\sigma$ (m\,s$^{-1}$) & Data Set\\
\hline
2455193.1992 & 70.86   & 2.97  & TLS \\
2455194.2968 & 69.48   & 3.31  & TLS \\
2455254.6953 & 117.42  & 5.17  & TLS \\
2455258.5781 & 64.17   & 4.84  & TLS \\
2455280.4375 & 69.01   & 3.13  & TLS \\
2455356.5078 & 44.13   & 3.94  & TLS \\
2455450.3476 & 31.79   & 3.49  & TLS \\
2455463.3515 & 29.63   & 2.41  & TLS \\
2455473.2578 & -10.70  & 2.38  & TLS \\
2455474.2421 & 38.88   & 2.47  & TLS \\
2455476.3281 & -55.21  & 1.70  & TLS \\
2455478.3203 & 50.83   & 1.80  & TLS \\
2455479.2421 & 79.10   & 2.54  & TLS \\
2455480.2460 & 7.88    & 2.35  & TLS \\
2455495.2343 & 59.36   & 2.50  & TLS \\
2455496.3515 & 31.56   & 2.51  & TLS \\
2455498.3515 & -9.58   & 2.56  & TLS \\
2455664.4609 & -76.03  & 3.09  & TLS \\
2455680.4062 & 48.53   & 4.90  & TLS \\
2455941.6953 & -37.13  & 6.54  & TLS \\
2456060.4882 & 24.67   & 13.26 & TLS \\
2456061.5039 & 22.32   & 12.00 & TLS \\
2456103.4531 & -42.60  & 9.56  & TLS \\
2456501.4687 & 46.14   & 3.06  & TLS \\
\hline
\end{tabular}
\caption{The RV data}
\end{center}
\label{tls}
\end{table}

\begin{table}
\begin{center}
\begin{tabular}{lc}
\hline
\hline
Parameter  & Value     \\
\hline
Period [days]                    & 702.47  $\pm$ 1.40            \\
T$_{0}$ [JD-2440000]             & 6610.16 $\pm$ 38.8           \\
$K$ [m\,s$^{-1}$]                & 148.4 $\pm$  4.1            \\
$e$                              & 0.08 $\pm$ 0.03            \\
$\omega$ [deg]                   & 234.26 $\pm$ 19.8           \\
$f(m)$ [solar masses]            & (2.35 $\pm$ 0.18) $\times$ 10$^{-7}$   \\
$m$ sin $i$ [$M_{Jupiter}$]      & 10.7 $\pm$ 0.6                           \\
$a$ [AU]                         &                               \\
\hline
\end{tabular}
\caption{Orbital parameters for the hypothetical companion to $\gamma$~Dra.
}
\end{center}
\label{orbitparm}
\end{table}

\begin{table}
\begin{center}
\begin{tabular}{ccc}
\hline
\hline
  Period          &    K                  & Phase      \\
	(d)       & (m\,s$^{-1}$)         &            \\
\hline
665.8 $\pm$ 3.8   &  125.73 $\pm$  6.95   & 0.32 $\pm$ 0.03 \\     
800.9 $\pm$ 10.2  &  45.52  $\pm$  7.17   & 0.97 $\pm$ 0.02 \\    
1854.9 $\pm$ 73.9 &  27.44  $\pm$  8.55   & 0.30 $\pm$ 0.11 \\   
\hline
\end{tabular}
\caption{Frequencies from pre-whitening of the RV data
}
\end{center}
\label{prewhite}
\end{table}

\end{document}